\documentclass[journal]{IEEEtran}
\usepackage{amsmath,amsfonts,amssymb,amsthm}
\usepackage{graphicx}
\usepackage{cite}
\usepackage{threeparttable}
\usepackage{booktabs}
\usepackage{mathrsfs}
\usepackage{multirow}
\usepackage{subfig}
\usepackage{bm}
\usepackage{float}
\usepackage{color}
\usepackage{xcolor}
\usepackage{makecell}
\usepackage{listings}
\usepackage[pass]{geometry}
\usepackage{enumitem}
\usepackage[linesnumbered,lined,boxed,vlined]{algorithm2e}
\usepackage[export]{adjustbox}
\usepackage{fancyhdr}
\usepackage{hyperref}
\SetArgSty{textup}
\SetAlFnt{\small}

\SetCommentSty{mycommfont}

\fancypagestyle{firstpage}{
  \fancyhead[C]{This paper is accepted by IEEE Transactions on Computer-Aided Design of Integrated Circuits and Systems (\url{https://ieeexplore.ieee.org/document/10767386})} 
  \fancyfoot[C]{}
}

\begin{document}

\title{CKTSO: High-Performance Parallel Sparse Linear Solver for General Circuit Simulations}
\author{Xiaoming Chen,~\IEEEmembership{Member,~IEEE}
\thanks{This work was supported by National Natural Science Foundation of China under Grant 62122076 and Grant 62488101.}
\thanks{X. Chen is with Institute of Computing Technology, Chinese Academy of Sciences, Beijing 100190, China (e-mail: chenxiaoming@ict.ac.cn).}
}


\maketitle
\thispagestyle{firstpage}

\begin{abstract}
This paper introduces CKTSO {(abbreviation of ``\underline{c}ir\underline{c}ui\underline{t} \underline{so}lver'')}, a novel sparse linear solver specially designed for the simulation  program with integrated circuit emphasis (SPICE). CKTSO is a parallel solver and can be run on a multi-core, shared-memory computer. The algorithms of CKTSO are designed by considering the features of matrices involved in SPICE simulations. CKTSO is superior to existing similar solvers mainly in the following three aspects. First, the matrix ordering step of CKTSO combines different types of ordering algorithms such that it can generally obtain the fewest fill-ins for a wide range of circuit matrices. Second, CKTSO provides a parallel fast LU factorization algorithm with pivot check, which behaves good performance, scalability, and numerical stability. Third, CKTSO provides a structure-adaptive hybrid parallel triangular solving algorithm, which can adapt to various circuit matrices. Experiments including both benchmark tests and SPICE simulations demonstrate the superior performance of CKTSO. The libraries of CKTSO are available at \url{https://github.com/chenxm1986/cktso}.
\end{abstract}


\section{Introduction}
\begin{figure}[t]
  \centering
  \includegraphics[width=.45\columnwidth]{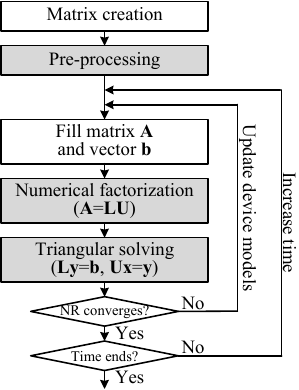}\\
  \caption{Sparse solver in a typical transient simulation flow.}\label{fig:tranflow}
\end{figure}

\IEEEPARstart{T}{he simulation} program with integrated circuit emphasis (SPICE)~\cite{spice} is a fundamental electronic design automation (EDA) kernel. SPICE utilizes numerical computation theories to find the detailed responses of integrated circuits. A SPICE simulation process involves solving a series of sparse linear systems. Fig.~\ref{fig:tranflow} illustrates a typical transient simulation flow. There are two nested levels of loops, where the outer loop iterates the time and a Newton-Raphson (NR) iteration is performed in each inner loop.
In each NR iteration, 
 a sparse linear system (${\bf Ax}={\bf b}$ where $\bf A$ is sparse) is solved.  The linear solver is usually the performance bottleneck of SPICE simulators. In post-layout simulations, the linear solver can spend more than half of the total simulation time~\cite{daniels_hspice_tech_2010}.

SPICE simulators generally use LU factorization~\cite{davis2016survey} to solve sparse linear systems, as linear systems in SPICE simulations are usually not well conditioned. 
 As shown in Fig.~\ref{fig:tranflow}, sparse LU factorization involves three main steps: \textit{pre-processing}, \textit{numerical factorization}, and \textit{triangular solving}.  Pre-processing  reorders the matrix to minimize fill-ins that will be generated during numerical factorization. It is performed only once if the matrix structure is fixed, which usually holds in SPICE simulations. In each NR iteration, the matrix is factorized into \textit{LU factors}: ${\bf A}={\bf LU}$, where $\bf L$ is a lower-triangular matrix and $\bf U$ is an upper-triangular matrix.  Finally, the solution $\bf x$ is obtained by solving two triangular systems, ${\bf Ly}\!=\!\bf b$ and ${\bf Ux}\!=\!\bf y$. Numerical factorization and triangular solving are executed in every NR iteration. Typically, numerical factorization spends more time than triangular solving.

{People have developed several LU factorization-based sparse solvers to accelerate circuit simulations. KLU~\cite{davis_klu_toms_2010}, developed in 2006, is a sequential implementation of the sparse left-looking algorithm~\cite{gilbert_gp_jssc_1988}. NICSLU~\cite{tcad13,tcasii}, developed in 2013, parallelizes the sparse left-looking algorithm at the column granularity. Later in 2016, Basker~\cite{basker} is developed, which partitions the matrix and parallelizes sparse LU factorization at the sub-matrix granularity. By exploring parallelism in the sparse left-looking algorithm, both NICSLU and Basker achieve speedups relative to KLU. PARDISO~\cite{pardiso}, which is a general-purpose sparse direct solver, has also shown good performance in semiconductor device simulations~\cite{pardiso2}.}

Despite the good performance of these solvers, it has been found that they have issues when facing some type of circuits. Various circuits put high demands for every step of LU factorization-based sparse linear solvers. \textbf{Challenge 1:} The popular matrix ordering methods, minimum degree and its variants~\cite{amd,mf,mmd}, which are adopted by KLU and NICSLU, usually do not perform well on post-layout, mesh-style, or very large circuits. Instead, nested dissection~\cite{nd} generally obtains better orderings on such matrices. However, nested dissection performs badly on pre-layout or small circuits. \textbf{Challenge 2:} The high sparsity of circuit matrices leads to low parallelism and a low computation-to-communication ratio in numerical LU factorization, which is difficult to be efficiently parallelized on multi-core processors. \textbf{Challenge 3:} For triangular solving, the parallelism and computation-to-communication ratio  are both extremely low, and it is extremely difficult to get speedups by parallelism. NICSLU provides parallel factorization but the solving step is sequential.

{My previous work proposed a fast numerical factorization algorithm~\cite{cktso} for circuit matrices, which addresses Challenge 2 and improves the performance and scalability of parallel LU factorization relative to NICSLU. This paper is extended from \cite{cktso} and introduces a complete solver, CKTSO (abbreviation of ``\underline{c}ir\underline{c}ui\underline{t} \underline{so}lver''), to accelerate circuit simulations. CKTSO is developed by redesigning the three steps of sparse solvers for SPICE simulators, to address all the above-mentioned challenges. CKTSO has the following three unique features.} \textcircled{1} The matrix ordering of CKTSO includes both minimum degree and nested dissection methods, and the best ordering is selected after trying them. \textcircled{2} For numerical factorization, CKTSO exploits the unique features of matrices in circuit simulation iterations and proposes a novel pivoting reduction-based fast LU factorization algorithm, which behaves good scalability and numerical stability.  \textcircled{3} For triangular solving, CKTSO proposes a novel structure-adaptive hybrid parallelism strategy. {Feature \textcircled{2} inherits from \cite{cktso} and features \textcircled{1} and \textcircled{3} are new relative to  \cite{cktso}. The three features are also the major technical advances of CKTSO relative to NICSLU.} The contributions of this work 
 are summarized as follows.
\begin{itemize}
\item A parallel fast LU factorization algorithm is proposed, which combines the advantages of both factorization with pivoting and re-factorization without pivoting, and behaves good performance and scalability, as well as good numerical stability.
\item A structure-adaptive  hybrid parallel triangular solving method is proposed. It adaptively splits the triangular matrices (i.e., $\bf L$ and $\bf U$) and uses different parallel triangular solving strategies, according to the characteristics of the symbolic structures of the sub-matrices.
\item Experimental results have revealed that CKTSO achieves better performance for a wide range of circuit matrices than stat-of-the-art solvers. In practical SPICE simulations, CKTSO also shows better performance in different simulation scenarios.
\end{itemize}

\section{Background}\label{sec:backg}

\subsection{Sparse Up-looking LU Factorization}

The sparse left-looking LU factorization~\cite{gilbert_gp_jssc_1988} is widely adopted by linear solvers for SPICE simulations (e.g., KLU~\cite{davis_klu_toms_2010},  NICSLU~\cite{tcad13,tcasii}, and Basker~\cite{basker}). It computes the LU factors column by column. CKTSO uses the row-major order, so the left-looking algorithm is transposed and can be named as sparse \textit{up-looking} LU factorization, as shown in Algorithm~\ref{algo:leftlooking}. It computes the LU factors row by row (line 1). As its name implies, when computing a row, some rows on its upper side will be ``looked at'' (used). When computing a row, symbolic prediction (line 2), numerical update (lines 3-5), and pivoting (lines 6-7) are executed. Since pivoting may change the structure of the LU factors, symbolic prediction cannot be decoupled from numerical computation. 

\begin{algorithm}[t]
\caption{Sparse up-looking LU factorization algorithm~\cite{gilbert_gp_jssc_1988}. If lines 2, 6, and 7 are removed, it becomes the up-looking LU re-factorization algorithm.}\label{algo:leftlooking}
\KwIn{$N\times N$ matrix $\bf A$ after pre-processing, pivoting threshold $\varepsilon$}
\KwOut{Matrices $\bf L$ and $\bf U$}
\For {$i=1:N$}
{
Symbolic prediction: find  symbolic structures of ${\bf L}(i,1\!:\!i)$ and ${\bf U}(i,i\!:\!N)$\;
${\bf x}={\bf A}(i,:)$\; \tcc{ $\bf x$ is a vector of length $N$ for holding intermediate results of current row}
\For {$j=1:i-1$ {where} ${\bf L}(i,j)$ { is nonzero}}
{
${\bf x}(j+1:N)-={\bf x}(j)\times {\bf U}(j,j+1:N)$\;
}
\If {$|{\bf x}(i)|<\varepsilon  \times \mathop {\max }\limits_{i + 1 \le k \le N} \left\{ {|{\bf{x}}(k)|} \right\}$}
{
Exchange ${\bf x}(i)$ with ${\bf x}(\mathop {\arg \max }\limits_{i + 1 \le k \le N} \left\{ {|{\bf{x}}(k)|} \right\})$\;
}
${\bf L}(i,1:i)={\bf x}(1:i)$\;
 ${\bf U}(i,i:N)={\bf x}(i:N)/{\bf x}(i)$\;
}
\end{algorithm}

 Symbolic prediction (line 2) determines the symbolic structure of the current row, based on the already computed rows on the upper side~\cite{gilbert_gp_jssc_1988,davis_klu_toms_2010}. 
 The symbolic structure of ${\bf L}(i,1:i-1)$  also determines which rows on the upper side will update the current row. Numerical update (lines 3-5) uses these dependent rows to update the values of the current row. 
 The core operation of numerical update is a series of floating-point multiply-and-accumulates (MACs) (line 5).  After that, pivoting (lines 6-7) is performed to ensure large diagonal elements. When the diagonal element is smaller than the maximum element in the current row of $\bf U$ times the pivoting threshold ($\varepsilon$), the diagonal element and the largest element in the current row of $\bf U$ are exchanged (line  7).
 
 The above described flow is a complete sparse up-looking LU factorization with pivoting. Before LU factorization, if the diagonal elements are assumed to be large enough, pivoting can be skipped. As a result, the symbolic structure of the LU factors will not change, and symbolic prediction can also be eliminated. In this case,  the algorithm becomes \textit{re-factorization} without pivoting, which only performs numerical update for each row. Re-factorization 
 reuses the symbolic structure and  pivoting order obtained in the last factorization with pivoting. If one or more small diagonal elements that do not meet the pivoting rule appear in re-factorization, they cannot be handled and may cause large errors in the solution.

\subsection{Parallel Sparse Up-looking LU Factorization}\label{sec:para}
To perform parallel up-looking LU factorization, dependencies between rows should be first determined. The sparsity of the LU factors implies the inter-row parallelism. Lines 4-5 of Algorithm~\ref{algo:leftlooking} indicate that row $i$ depends on row $j$ (i.e., row $i$ needs row $j$'s values for numerical update) if and only if ${\bf L}(i,j)$ is a nonzero element. 
In short, the symbolic structure of $\bf L$  determines the inter-row dependencies. Based on this principle, a dependency graph can be built based on the symbolic structure of $\bf L$. There is an edge $j\to i$ in the dependency graph if ${\bf L}(i,j)$ is a nonzero element. 
This dependency graph can be named as an \textit{elimination graph (EGraph)}. The EGraph describes exact inter-row dependencies corresponding to a specific symbolic structure of the LU factors, and can be used to schedule parallel LU re-factorization without pivoting~\cite{tcasii}.

For parallel factorization with pivoting, however, it is impossible to determine the exact inter-row dependencies prior to factorization, since the symbolic structure of $\bf L$ depends on pivoting and is known only after factorization is completed. To solve this dilemma, people have proposed a concept of \textit{elimination tree (ETree)}~\cite{liu_etree_jmaa_1990}, which describes an \textbf{upper bound} of the inter-row dependencies when considering pivoting. The meaning of the upper bound is that, the dependencies generated by \textbf{any} pivoting order of up-looking factorization are contained in the ETree.  The ETree is constructed from the symbolic structure of matrix $\bf A$ {(if $\bf A$ is unsymmetric, the ETree is built from ${\bf A}^\mathsf{T}{\bf A}$ without explicitly forming ${\bf A}^\mathsf{T}{\bf A}$~\cite{liu_etree_jmaa_1990})}, so it can be built in the pre-processing step.

\begin{figure}[t]
  \centering
  \includegraphics[width=.95\columnwidth]{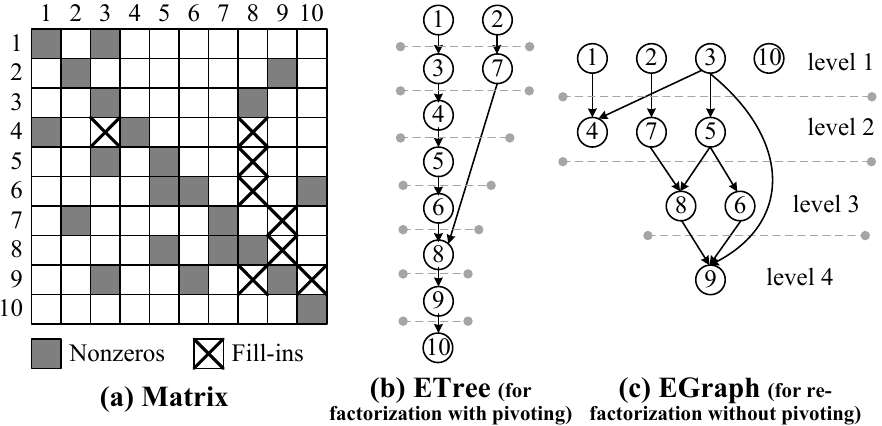}\\
  \caption{Matrix example and its corresponding ETree and EGraph (assuming that the pivots are the original diagonal elements).}\label{fig:example}
\end{figure}

\begin{figure*}[t]
  \centering
  \includegraphics[width=.9\textwidth]{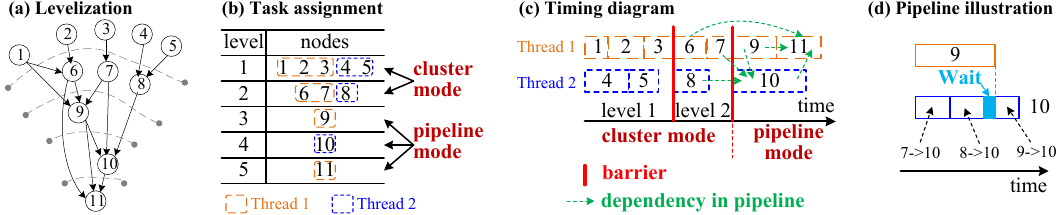}\\
  \caption{Levelization-based dual-mode parallel scheduling method~\cite{tcad13,tcasii}. This example does not correspond to Fig.~\ref{fig:example}.}\label{fig:para}
\end{figure*}

The EGraph and ETree are both directed acyclic graphs, as illustrated in Fig.~\ref{fig:example}. The parallelism implied in EGraph and ETree can be explored by levelizing them, so that parallel factorization and re-factorization can both be scheduled by a unified method~\cite{tcad13,tcasii}. As shown in Fig.~\ref{fig:para}, the EGraph or ETree can be levelized (Fig.~\ref{fig:para}(a)) so that  nodes (a node corresponds to a row) in the same level have no dependency. The \textit{level} of a node is the maximum path length from any source node to itself. A dual-mode scheduling method has been proposed~\cite{tcad13,tcasii}. For front levels that have a large number of nodes in each level, they are computed level by level. Nodes in a level are evenly assigned to threads and computed in parallel. This method is called a \textit{cluster mode}. For the remaining levels that have very few nodes in each level, a \textit{pipeline mode} is proposed which explores finer-grained parallelism between dependent rows. Fig.~\ref{fig:para}(d) illustrates an example of computing node 10 using the pipeline mode, corresponding to the dependencies described in Fig.~\ref{fig:para}(a). Nodes 9 and 10 are computed in parallel in the pipeline mode. Since node 10 depends on node 9, node 10 can first use already-finished nodes (i.e., nodes 7 and 8) to perform partial updates. These partial updates are performed in parallel with the computation of node 9. This is the key to explore parallelism between dependent rows. After node 9 is finished, node 10 can use node 9's results to perform a partial update. The levelization-based dual-mode scheduling method has been adopted by NICSLU.

The height and width of the ETree/EGraph can be an intuitive estimation of the scalability of parallel LU factorization/re-factorization. The height (maximum level) is the critical path length and the width  (maximum number of nodes in each level) is the maximum parallelism. As illustrated in Fig.~\ref{fig:example}, the ETree is tall and narrow, while the EGraph is short and wide. As the ETree describes an inter-row dependency upper bound, it implies many redundant dependencies. We have tested more than 50 circuit matrices from the SuiteSparse Matrix Collection~\cite{davis_matrix_web}. On average, the ETree is nearly 80$\times$ taller than the EGraph, while the EGraph is 10$\times$ wider. Therefore, the scalability of parallel factorization with pivoting (scheduled by the ETree) is much poorer than that of re-factorization without pivoting (scheduled by the EGraph). According to the results in Ref.~\cite{book}, for NICSLU, 16-threaded factorization with pivoting achieves only 2$\times$ speedup on average, compared with sequential factorization.

\subsection{Triangular Solving}\label{sec:back_solve}
\begin{algorithm}[t]
\caption{Lower triangular solving of ${\bf Ly}=\bf b$.}\label{algo:solve}
\KwIn{Matrix $\bf L$ obtained from LU factorization, vector $\bf b$}
\KwOut{Vector $\bf y$}
${\bf y}={\bf b}$\;
\For {$i=1:N$}
{
$s=0$\;
\For {$j=1:i-1$ where ${\bf L}(i,j)$ is nonzero}
{
$s+={\bf y}(j)\times {\bf L}(i,j)$\;
}
${\bf y}(i)=({\bf y}(i)-s)/{\bf L}(i,i)$\;
}
\end{algorithm}

Algorithm~\ref{algo:solve} shows the lower triangular solving of ${\bf Ly}=\bf b$. With the row-major order, solving is also performed in a row-by-row manner (line 2). For each row, it traverses the nonzero elements of the current row of $\bf L$ (line 4), performs a MAC operation for each nonzero element (line 5), and finally computes an element of the solution vector (line 6). The upper triangular solving of ${\bf Ux}=\bf y$ is similar, with the only difference that $\bf U$ is traversed in the reversed row order.

For solving ${\bf Ly}=\bf b$, it can easily be seen that, if ${\bf L}(i,j)$ ($j<i$) is a nonzero element, the solving of ${\bf y}(i)$ depends on  ${\bf y}(j)$. Based on this principle, one may also build a dependency graph to parallelize triangular solving. However, unlike LU factorization and re-factorization, triangular solving involves an extremely low computation-to-communication ratio. In fact, each nonzero element of $\bf L$ and $\bf U$ participates in exactly one MAC operation, and the primary bottleneck of triangular solving is not computation but memory access. As a result, parallelization is extremely difficult, since the memory access and inter-thread communication overheads caused by parallelization can be much larger than the computational workload.

\subsection{Related Works}
There are some popular general-purpose sparse solvers (e.g., SuperLU~\cite{slu}, PARDISO~\cite{pardiso},  UMFPACK~\cite{davis_umfpack_tms_2004}, et al.).  In recent years, KLU~\cite{davis_klu_toms_2010}, NICSLU~\cite{tcasii,tcad13}, {and Basker~\cite{basker}} are proposed to accelerate circuit simulations, which are based on the sparse left-looking algorithm~\cite{gilbert_gp_jssc_1988}. The primary difference between the design principles of these circuit simulation-oriented solvers and general-purpose sparse solvers is due to the matrix sparsity. As matrices in circuit simulations are generally much sparser than  matrices from other applications~\cite{davis_klu_toms_2010}, solvers designed for circuit simulations typically do not assemble nonzero elements to form big dense blocks which can be accelerated by dense linear algebra kernels. KLU~\cite{davis_klu_toms_2010} is sequential while NICSLU~\cite{tcasii,tcad13} provides parallel left-looking LU factorization and re-factorization. Factorization with pivoting and re-factorization without pivoting of NICSLU are scheduled by the ETree and EGraph, respectively. It is reported that NICSLU  achieves better performance than other popular solvers (KLU, PARDISO, SuperLU, UMFPACK, etc.) in real circuit and power grid simulation problems~\cite{power_pes2015,power_2019,power2015}. However, the scalability of NICSLU's factorization with pivoting is not so good (2$\times$ speedup using 8-16 threads)~\cite{book}. {Basker~\cite{basker} parallelizes LU factorization based on matrix partitioning. It partitions the matrix to a block triangular form where each diagonal block is further partitioned using nested dissection, generating a number of independent sub-matrices. It performs well on highly sparse matrices but does poorly on high fill-in density matrices (e.g., post-layout matrices).}

For triangular solving, current sparse solvers designed for circuit simulations mostly use sequential solving, including NICSLU which provides parallel factorization and re-factorization. Parallel triangular solving in general-purpose sparse solvers (e.g., \cite{bollhofer2020state,solve2}) is typically based on some dependency analysis method which is similar to that used for parallel LU factorization.  However, this method is inefficient for highly-sparse circuit matrices, since there is little parallelizable workload and the scheduling overhead may be larger than the computational workload. {Ref.~\cite{psolve} uses level scheduling and recursive blocking to parallelize triangular solving. The method behaves well for general matrices but may not be suitable for highly-sparse circuit matrices as the blocking scheme is too fine grained which causes heavy scheduling overhead.} Ref.~\cite{solve} uses a nested dissection-based partitioning method to parallelize triangular solving for power grid simulations, which is not for general circuit simulations.

\section{CKTSO Overview}

\begin{figure}[t]
  \centering
  \includegraphics[width=.75\columnwidth]{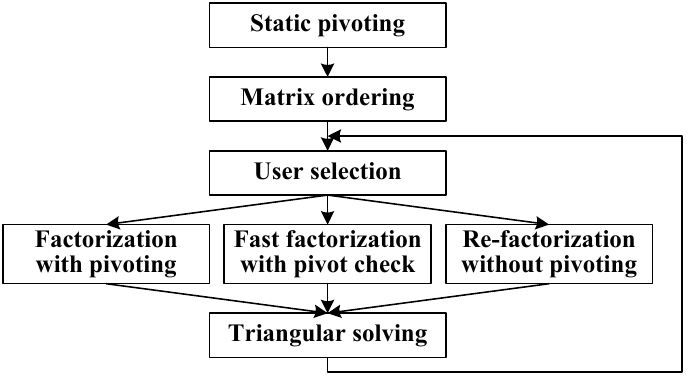}\\
  \caption{Overall flow of CKTSO.}\label{fig:flow}
\end{figure}

\subsection{Pre-processing}

Fig.~\ref{fig:flow} shows the overall flow of CKTSO. For pre-processing, {CKTSO first performs a static pivoting step based on the maximum weighted matching algorithm~\cite{staticpiv,staticpiv2}. The maximum weighted matching algorithm} permutes large elements to the diagonal, such that the absolute value of the product of the diagonal elements is maximized. It greatly reduces the possibility of dynamic pivoting during numerical factorization (but dynamic pivoting cannot be completely avoided via static pivoting). It also calculates row and column scaling vectors that can scale the matrix values, such that the diagonal elements are 1 or -1, and non-diagonal elements are bounded in [-1,1]. 
The scaling feature of CKTSO is controlled by users. 
 Mathematically, static pivoting is to find a row permutation matrix $\bf P$ and two diagonal scaling matrices $\bf S_r$ and $\bf S_c$. The matrix after static pivoting is ${\bf A'}=\bf PA$ (without scaling) or ${\bf A'}=\bf S_rPAS_c$ (with scaling).

\begin{figure}[t]
  \centering
  \includegraphics[width=.7\columnwidth]{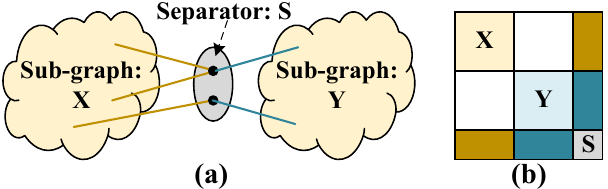}\\
  \caption{Nested dissection ordering. (a) Graph partitioning. (b) Corresponding bordered block diagonal matrix.}\label{fig:nd}
\end{figure}

After static pivoting, the matrix is reordered to minimize fill-ins that will be generated during numerical factorization. Mathematically, matrix ordering is to find a symmetric permutation matrix $\bf Q$, such that factorization of $\bf A''= QA'Q^\mathsf{T}$ will generate the minimum fill-ins. Finding an ordering with the minimum fill-ins is an NP-complete problem~\cite{np}, so heuristics are used in practice.  Minimum degree~\cite{amd,mf,mmd} and nested dissection~\cite{nd} are two popular kinds of matrix ordering methods. Minimum degree is performed in an iterative manner and selects a node with the minimum degree in each iteration to be the pivot, and forms new fill-ins by eliminating the pivot. It is adopted by KLU~\cite{davis_klu_toms_2010} and NICSLU~\cite{tcasii,tcad13}, and generally performs well for small- to medium-scale matrices (e.g., pre-layout circuits). For large-scale matrices or matrices of post-layout circuits,  nested dissection usually generates better orderings. 
One-level nested dissection partitions a graph which represents a matrix into three parts, a node separator whose size is minimized and two sub-graphs (Fig.~\ref{fig:nd}(a)), such that removal of the node separator disconnects the two sub-graphs. If the node separator is ordered last (Fig.~\ref{fig:nd}(b)), no fill-in will occur between nodes of the two sub-graphs. 
Each sub-graph can be recursively ordered by the same method.    Nested dissection is adopted by PARDISO~\cite{pardiso}.

Since minimum degree and nested dissection respectively perform well on different types of matrices, CKTSO uses a combination of them. For minimum degree, CKTSO uses the popular approximate minimum degree algorithm~\cite{amd} and its variant, the approximate minimum deficiency algorithm~\cite{mf}. For nested dissection, though METIS~\cite{metis} provides a widely-used implementation, it has some issues. METIS 
recursively partitions a graph, until the sub-graphs' sizes are smaller than a threshold. Then, each sub-graph is ordered by the multiple minimum degree algorithm~\cite{mmd} independently, and the node separator uses the natural order without ordering.  By using Fig.~\ref{fig:nd}(b) to explain the issues, though the orderings of $\bf X$ and $\bf Y$ will not impact each other, they will impact the fill-ins of $\bf S$ through the two boundaries. This implies that, when ordering $\bf X$ and $\bf Y$, their impacts on the fill-ins of $\bf S$ should be taken into account. This can be realized by applying a constrained minimum degree algorithm~\cite{cmd} after the incomplete nested dissection. The constrained minimum degree orders the entire matrix, where each pivot has a constraint of its group index and the pivots must be eliminated in the order of group. For example, in Fig.~\ref{fig:nd}, the pivots in $\bf X$, $\bf Y$, and $\bf S$ are in groups 0, 1, and 2, respectively. By using the group to constrain the pivot elimination order, this method guarantees the sub-matrix-level order returned by the incomplete nested dissection, while all  diagonal sub-matrices are properly ordered. CKTSO integrates a modified METIS (by removing the independent minimum degree ordering for each diagonal sub-matrix) followed by the constrained approximate minimum degree algorithm~\cite{cmd}.

In the matrix ordering phase, CKTSO runs all the integrated ordering methods in parallel and selects the best with the minimum fill-ins. The combined ordering method can generally obtain the fewest fill-ins for a wide range of circuit matrices, and thus, addresses the challenge that a single ordering method may not be good for various circuit matrices. According to experiments, the pre-processing time of CKTSO is on average about 3$\times$ of that of NICSLU, and similar to that of PARDISO. Since pre-processing is a one-time procedure in a SPICE simulation, the pre-processing overhead is not so significant and can be hidden by the large number of more time-consuming NR iterations. 

\subsection{Numerical Factorization and Triangular Solving}
CKTSO provides three LU factorization functions: factorization with pivoting, re-factorization without pivoting, and fast factorization with pivot check. They can be selected by users according to the scenario. The basic principles of the former two functions as well as their parallelization methodology are introduced in Section~\ref{sec:backg}. In fact, the factorization and re-factorization functions of CKTSO are same as those of NICSLU~\cite{tcasii,tcad13}, so they will not be introduced in this paper. CKTSO provides a novel parallel fast factorization algorithm by skipping pivoting but the numerical stability is maintained by pivot check. This algorithm will be introduced in Section~\ref{sec:factor}. CKTSO also provides a structure-adaptive  hybrid parallel triangular solving algorithm, which will be introduced in Section~\ref{sec:solve}. In the rest of the paper, the notation of $\bf A$ will be used to denote the matrix after pre-processing.

\section{Parallel Fast Factorization with Pivot Check}\label{sec:factor}

{This section introduces the pivot check-based parallel fast LU factorization algorithm~\cite{cktso} of CKTSO, which aims to address the scalability challenge of parallel LU factorization of NICSLU~\cite{tcad13,tcasii}. }


The re-factorization without pivoting provided by KLU and NICSLU reuses the pivoting order and the symbolic structure of the LU factors obtained from the last factorization with pivoting, so symbolic-related operations can be eliminated. This is numerically unstable since the matrix values are changing during circuit simulation iterations, and 
disabling pivoting may generate erroneous solutions. To ensure correct solutions, factorization and re-factorization must be carefully selected in each circuit simulation iteration. However, this is not an easy task before numerical factorization. 
Evaluating whether the values of matrix $\bf A$ change much cannot give a stable prediction, as a small change in a few sensitive elements may change the LU factors' values dramatically.  

 In circuit simulations, 
  when the NR iterations of an operating-point (OP) simulation or at each time node in a transient simulation are near convergence, the matrix values tend to change smoothly, and successive NR iterations tend to reuse the previous pivoting order.    This intuition provides an opportunity to ``guess'' the task dependencies before numerical factorization. If the structure of the LU factors and the pivoting order are aggressively assumed to be unchanged from the previous factorization, a ``guessed'' EGraph can be tentatively used to schedule parallel fast factorization, skipping symbolic prediction and pivoting, but the pivots should be checked.  Based on this idea, CKTSO employs a novel parallel fast LU factorization algorithm which explores the advantages of both factorization with pivoting and re-factorization without pivoting. In most circuit simulation iterations, its performance and scalability are similar to those of re-factorization but the numerical stability is similar to that of factorization. 

Since the EGraph is ``guessed'' based on the structure of the LU factors obtained in the last  factorization with pivoting, when an unsatisfactory pivot is detected, re-pivoting is needed and the structure of the remaining LU factors will change. The EGraph will also change so it  can no longer be used. Instead, the ETree that implies an upper bound of the dependencies will be used to schedule the remaining factorization with pivoting. A key question raises here: which nodes will be computed with pivoting? The answer is determined by the ETree. In fact, some finished nodes may also need to be re-computed. For instance, in Fig.~\ref{fig:example}(c), if the first two levels of the EGraph are finished and an unsatisfactory pivot is found in node 6, nodes 6, 8, 9, and 10 need to be computed with pivoting according to the ETree shown in Fig.~\ref{fig:example}(b), although nodes 8 and 10 are finished. In circuit simulation iterations, since the matrix values usually change smoothly, especially when the NR iterations are near convergence,  the ``guessed'' EGraph is highly likely a correct prediction and re-pivoting tends not to occur. 
By using this strategy, the advantages of both factorization with pivoting and re-factorization without pivoting are given full play. 
Note that the original re-factorization provided by KLU and NICSLU does not check pivots. Even if pivot check is added, they can only exit when an unsatisfactory pivot is found. Instead, CKTSO provides a novel mechanism to restart factorization with pivoting. This is the primary advantage of the proposed fast LU factorization relative to re-factorization.

\subsection{Parallel Fast Factorization Flow}

\begin{figure}[t]
  \centering
  \includegraphics[width=.9\columnwidth]{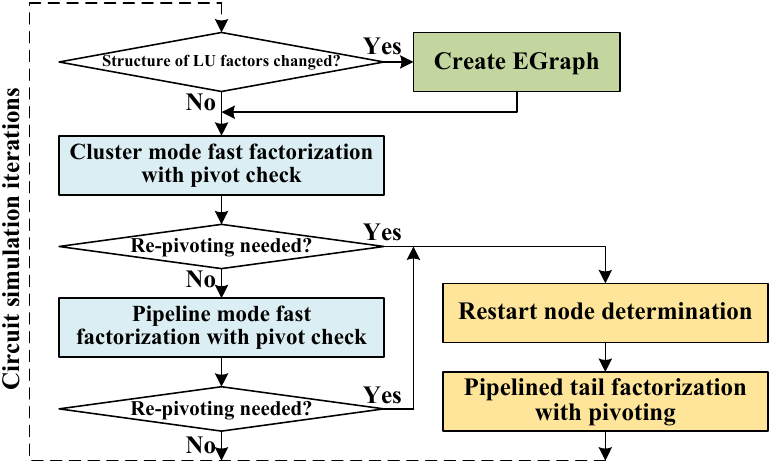}\\
  \caption{Flow of parallel fast LU factorization algorithm.}\label{fig:factor}
\end{figure}

Fig.~\ref{fig:factor} shows the flow of the proposed parallel fast LU factorization algorithm. 
A ``guessed'' EGraph is first built from the already obtained structure of the LU factors. The ``guessed'' EGraph will not be re-built in successive NR iterations, unless the symbolic structure of the LU factors changes. For each factorization, parallel fast factorization with pivot check is tentatively scheduled based on the ``guessed'' EGraph. If all pivots are satisfactory, the structure of the LU factors does not change and the ``guessed'' EGraph is correct. This is the best case with the entire matrix computed by parallel fast factorization with pivot check, which has similar performance and scalability to re-factorization without pivoting. If one or more unsatisfactory pivots are detected, re-pivoting is needed and  fast factorization with pivot check is interrupted. A pipelined tail factorization with pivoting, scheduled by the ETree, is then executed to compute the remaining matrix. Before that, the row from which tail factorization with pivoting starts is determined. The worst case happens when the first row needs re-pivoting so the entire matrix is computed with pivoting, which has similar performance and scalability to factorization with pivoting.

\subsection{Parallel Fast Factorization with Pivot Check}
CKTSO also employs the levelization-based methodology~\cite{tcasii,tcad13} to explore parallelism in the EGraph and ETree. The ``guessed'' EGraph is levelized by calculating the levels of all nodes, as illustrated in Fig.~\ref{fig:para}(a). Nodes in the same level can be computed in parallel without any dependency. A levelized EGraph generally looks like a funnel, as shown in Fig.~\ref{fig:example}(c). There is a common feature that the number of nodes in a level tends to become fewer with level increasing. Like NICSLU~\cite{tcasii,tcad13}, based on the levelized ``guessed'' EGraph, CKTSO also uses a cluster mode and a pipeline mode to schedule parallel fast factorization with pivot check,  as shown in Figs.~\ref{fig:para}(b) and \ref{fig:para}(c). A dividing level to distinguished the two modes is defined as the first level in EGraph which has less than \#threads$\times\alpha$ ($\alpha$ is an empirical parameter and is 2.0 in CKTSO) nodes.  Levels before and after the dividing level are parallelized using  the cluster mode and pipeline mode, respectively. Algorithms~\ref{algo:cluster} and \ref{algo:pipeline} show the cluster mode and pipeline mode of fast factorization with pivot check, respectively.

\subsubsection{Cluster Mode}

\begin{algorithm}[t]
\caption{Cluster mode of fast factorization with pivot check.}\label{algo:cluster}
$interrupted=0$; \tcc{shared variable for all threads}
\For{$L=1:L_{Cluster}$}
{
\For{threads \textit{in parallel}}
{
\For {row $i$ assigned to this thread}
{
${\bf x}={\bf A}(i,:)$\;
\For {$j=1:i-1$ {where} ${\bf L}(i,j)$ { is nonzero}}
{
${\bf x}(j+1\!:\!N)-={\bf x}(j)\times {\bf U}(j,j+1\!:\!N)$\;
}
\If {$|{\bf x}(i)|<\varepsilon  \times \mathop {\max }\limits_{i + 1 \le k \le N} \left\{ {|{\bf{x}}(k)|} \right\}$}
{
$interrupted=1$\;
\textbf{continue}\;
}
${\bf L}(i,1:i)={\bf x}(1:i)$\;
 ${\bf U}(i,i:N)={\bf x}(i:N)/{\bf x}(i)$\;
 $ finish[i]=1$\;
}
}
\textbf{barrier}\;
\If{$interrupted$}
{
\textbf{exit thread}\;
}
}
\end{algorithm}

In the cluster mode  of fast factorization with pivot check (Algorithm~\ref{algo:cluster}), the ``guessed'' EGraph is processed level by level (line 2). For each level, the nodes are evenly assigned to the threads and different threads compute the assigned nodes in parallel (line 3). For a specific node (row), numerical update (lines 5-7) is first performed just like re-factorization. Then, the pivot (i.e., the diagonal element) is checked (line 8). If the pivot  is smaller than the maximum element in the current row of $\bf U$ times the pivoting threshold $\varepsilon$, it does not conform to the pivoting requirement (line 8). To tell the other threads that an unsatisfactory pivot has been found, a shared variable $interrupted$ is set to 1 (line 9). 
After the computation of a level is completed, all threads are synchronized through a barrier (line 14). After that, if $interrupted$ is 1,  all threads will exit (lines 15-16); otherwise they will continue to compute the next level.  If the cluster mode finishes without detecting any unsatisfactory pivot, the pipeline mode of fast factorization with pivot check will be executed to compute the remaining rows.

\subsubsection{Pipeline Mode}\label{sec:pipe}

\begin{algorithm}[t]
\caption{Pipeline mode of fast factorization with pivot check.}\label{algo:pipeline}
$interrupted=0$; \tcc{shared variable for all threads}
\For{threads \textit{in parallel}}
{
\For {row $i$ assigned to this thread}
{
${\bf x}={\bf A}(i,:)$\;
\For {$j=1:i-1$ {where} ${\bf L}(i,j)$ { is nonzero}}
{
\While {$!finish[j]$}
{
\If{$interrupted$}
{
\textbf{exit thread}\;
}
}
${\bf x}(j+1:N)-={\bf x}(j)\times {\bf U}(j,j+1:N)$\;
}
\If {$|{\bf x}(i)|<\varepsilon  \times \mathop {\max }\limits_{i + 1 \le k \le N} \left\{ {|{\bf{x}}(k)|} \right\}$}
{
$interrupted=1$\;
\textbf{exit thread} \;
}
${\bf L}(i,1:i)={\bf x}(1:i)$\;
 ${\bf U}(i,i:N)={\bf x}(i:N)/{\bf x}(i)$\;
 $ finish[i]=1$\;
}
}
\end{algorithm}

There are very few nodes in each level of the tail part of the ``guessed'' EGraph,  and they usually look like a sequential chain. Hence, there is little inter-node parallelism and the cluster mode is inefficient for these highly-dependent nodes. Instead, the pipeline mode of fast factorization with pivot check (Algorithm~\ref{algo:pipeline}), which explores parallelism between dependent nodes, is used for such tasks.  The nodes belonging to the pipeline mode are first arranged to a sequence, and they are assigned to the threads on-the-fly. A pointer {\tt max\_busy} is maintained which points to the maximum position in the node sequence which is being computed. If a thread needs to get a new node to compute, it atomically increases {\tt max\_busy} by 1 and fetches the resulting value. The atomic operation guarantees that no two threads will get the same node.
 All threads compute the fetched nodes in parallel in an asynchronous way.   The pipeline mode overlaps the computation of dependent rows. 

In the pipeline mode, when numerically updating a row (lines 4-9), some dependent predecessors may be unfinished, but the finished predecessors can be used immediately. Each thread waits for any dependent predecessor to finish (line 6) before using it. After all predecessors are finished and have been used to update the current row, the pivot is checked (line 10). If the pivot is unsatisfactory, the shared variable, $interrupted$, is set to 1 (line 11), to tell other threads, and the current thread will exit immediately (line 12). This may cause a deadlock problem which happens if a thread has exited due to an unsatisfactory pivot and another thread is waiting for it. To avoid deadlock, when a thread is waiting for a dependent predecessor, $interrupted$ must also be checked (lines 6-8).

\subsection{Restart Node Determination}

\begin{figure}[t]
  \centering
  \includegraphics[width=.38\columnwidth]{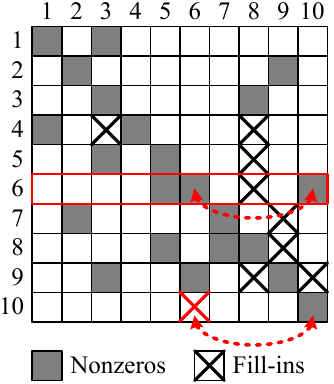}\\
  \caption{Example of re-pivoting on row 6: columns 6 and 10 are exchanged, incurring an additional dependency of row 10.}\label{fig:example2}
\end{figure}

During parallel fast factorization with pivot check, if some row incurs an unsatisfactory pivot,  re-pivoting is needed for that row, which will change the structure of the LU factors and make the dependencies described by the ``guessed'' EGraph partially incorrect. Hence, the remaining matrix must be computed with pivoting. Furthermore, due to the change of the dependencies, some already finished rows also need to be re-computed with pivoting. The matrix example shown in Fig.~\ref{fig:example}(a) is used to illustrate why this could happen, whose corresponding EGraph is in Fig.~\ref{fig:example}(c). In the EGraph, assume that the first two levels are finished and node 6 incurs an unsatisfactory pivot and will be re-pivoted.  Assume that columns 6 and 10 are exchanged due to re-pivoting of row 6, as illustrated in Fig.~\ref{fig:example2}. This column exchange introduces an additional nonzero fill-in at ${\bf L}(10,6)$, which brings an additional dependency of 6$\to$10. However, as shown in Fig.~\ref{fig:example}(c), this dependency does not exist in the EGraph. This is why row 10 is already finished in fast factorization with pivot check which is scheduled by the ``guessed'' EGraph. However, if re-pivoting happens on row 6 like this example, the computed result of row 10 becomes incorrect since the dependency of row 10 is changed.


Since the ETree describes an upper bound of the dependencies for any pivoting order, the ETree can determine the rows which will be re-computed. In the above-mentioned example, it can easily be derived from the ETree in Fig.~\ref{fig:example}(b) that rows 6, 8, 9, and 10 will be computed with pivoting. 
The general method to determine which nodes will be computed with pivoting is simple. All unfinished nodes in the ``guessed'' EGraph are traversed, and for each unfinished node, all of its descendants in the ETree need to be computed with pivoting.  All such nodes are put into a topologically-ordered sequence and computed by the pipelined tail factorization with pivoting.


\subsection{Pipelined Tail Factorization with Pivoting}

\begin{algorithm}[t]
\caption{Pipelined tail factorization with pivoting.}\label{algo:tail}
\For {threads \textit{in parallel}}
{
\For {row $i$ assigned to this thread}
{
	$\mathcal{U}=\emptyset$\;
	${\bf x}={\bf A}(i,:)$\;
	\tcc{{pre-factorization}}
	\While {predecessors of $i$ not all finished}
	{
		$\mathcal{F}=\emptyset$\;
		Symbolic prediction, in which unfinished predecessors are skipped, and put detected dependent rows that are not in $\mathcal{U}$ into $\mathcal{F}$\;
		\For {$j\in \mathcal{F}$}
{
${\bf x}(j+1\!:\!N)-={\bf x}(j)\times {\bf U}(j,j+1\!:\!N)$\;
}
$\mathcal{U}=\mathcal{U}\cup \mathcal{F}$\;
	}
	\tcc{{post-factorization}}
	Symbolic prediction: find symbolic structures of ${\bf L}(i,1:i)$ and ${\bf U}(i,i:N)$\;
\For {$j=1:i-1$ {where} ${\bf L}(i,j)$ { is nonzero} and $j\notin \mathcal{U}$}
{
${\bf x}(j+1:N)-={\bf x}(j)\times {\bf U}(j,j+1:N)$\;
}

\If {$|{\bf x}(i)|<\varepsilon  \times \mathop {\max }\limits_{i + 1 \le k \le N} \left\{ {|{\bf{x}}(k)|} \right\}$}
{
Exchange ${\bf x}(i)$ with ${\bf x}(\mathop {\arg \max }\limits_{i + 1 \le k \le N} \left\{ {|{\bf{x}}(k)|} \right\})$\;
}
${\bf L}(i,1:i)={\bf x}(1:i)$\;
 ${\bf U}(i,i:N)={\bf x}(i:N)/{\bf x}(i)$\;
 $ finish[i]=1$\;
}
}
\end{algorithm}

If re-pivoting happens, the remaining matrix (including those finished rows which will be re-computed with pivoting) will be computed using a pipelined tail factorization algorithm with pivoting. The tail factorization only has the pipeline mode, because  the ETree is tall and narrow, and  there are usually very few rows that can be computed using the cluster mode. Algorithm~\ref{algo:tail} shows the pipeline mode factorization with pivoting. The same dynamic scheduling method by utilizing atomic operations is used to assign rows to threads on-the-fly, as introduced in Section~\ref{sec:pipe}. The pipelined factorization flow of each thread can be divided into two parts: pre-factorization (lines 3-10) and post-factorization (lines 11-18). 

{The pre-factorization is the key to explore parallelism between dependent rows, which is also the implication of ``pipeline''. In pre-factorization, the already finished, dependent predecessors of the computing row ($i$) will update row $i$, before all predecessors of row $i$ are finished.  The pre-factorization will not end until all predecessors of row $i$ are finished. In the {\tt while} loop (lines 5-10), the finished predecessors are used to update row $i$, while those unfinished predecessors are skipped in both symbolic prediction and numerical update. Set $\mathcal{F}$ maintains the newly detected finished predecessors that will update row $i$, and set $\mathcal{U}$ records the rows that have already been used by row $i$.} 
{After all predecessors of row $i$ are finished, it enters post-factorization}. A complete symbolic prediction (line 11) without skipping any predecessors is first performed to determine the complete symbolic structure of row $i$. Then, those skipped predecessors in pre-factorization (i.e., the dependent rows which are not in $\mathcal{U}$) are now used to update row $i$ (lines 12-13). After numerical update, pivoting (lines 14-15) is performed.

\section{Parallel Triangular Solving}\label{sec:solve}

{This section introduces the structure-adaptive hybrid parallel triangular solving algorithm of CKTSO. This is one of the major technical advances of CKTSO over NICSLU~\cite{tcad13,tcasii} and \cite{cktso}, aiming at addressing the challenge of extremely low parallelism and computation-to-communication ratio of triangular solving of circuit matrices.}

Although numerical factorization spends longer time than triangular solving, triangular solving may also become the performance bottleneck in SPICE simulators. For example, in a linear circuit simulation with a fixed time step, only one factorization step is needed but a number of triangular solving steps are performed. Triangular solving is extremely difficult to parallelize, because triangular solving is highly sequential and the computational workload is too small. As mentioned in Section~\ref{sec:back_solve}, each nonzero element in $\bf L$ and $\bf U$ only participates in one MAC operation. CPU threads are not suitable for scheduling such fine-grained computations, as the scheduling overhead will be much higher than the computational cost. To minimize the scheduling overhead, one should group the elements with some similar features such that they can be scheduled together.

Problem partitioning is usually the key to explore parallelism and also to reduce parallel overhead. How to partition the triangular matrices becomes the most important challenge for parallelizing triangular solving. Using a single partitioning method for an entire triangular matrix (e.g., the method of Ref.~\cite{solve}) may not be the best solution, because the nonzero elements in $\bf L$ and $\bf U$ are not distributed uniformly. Instead, their distribution has obvious features.  The nonzero elements usually become denser on the right-bottom corner. This is due to the nature of LU factorization. This feature provides a possibility to partition the LU factors using different methods according to the nonzero element distribution so that different parallelism strategies can be used for different sub-matrices, according to their characteristics of the nonzero structure. By scheduling tasks at the sub-matrix level, the structure-adaptive method also reduces the scheduling overhead. In this section, the lower triangular solving will be used as an example to illustrate the parallel triangular solving methodology of CKTSO, and the upper triangular solving method is similar.

\begin{figure}[t]
  \centering
  \includegraphics[width=.55\columnwidth,frame=.01pt]{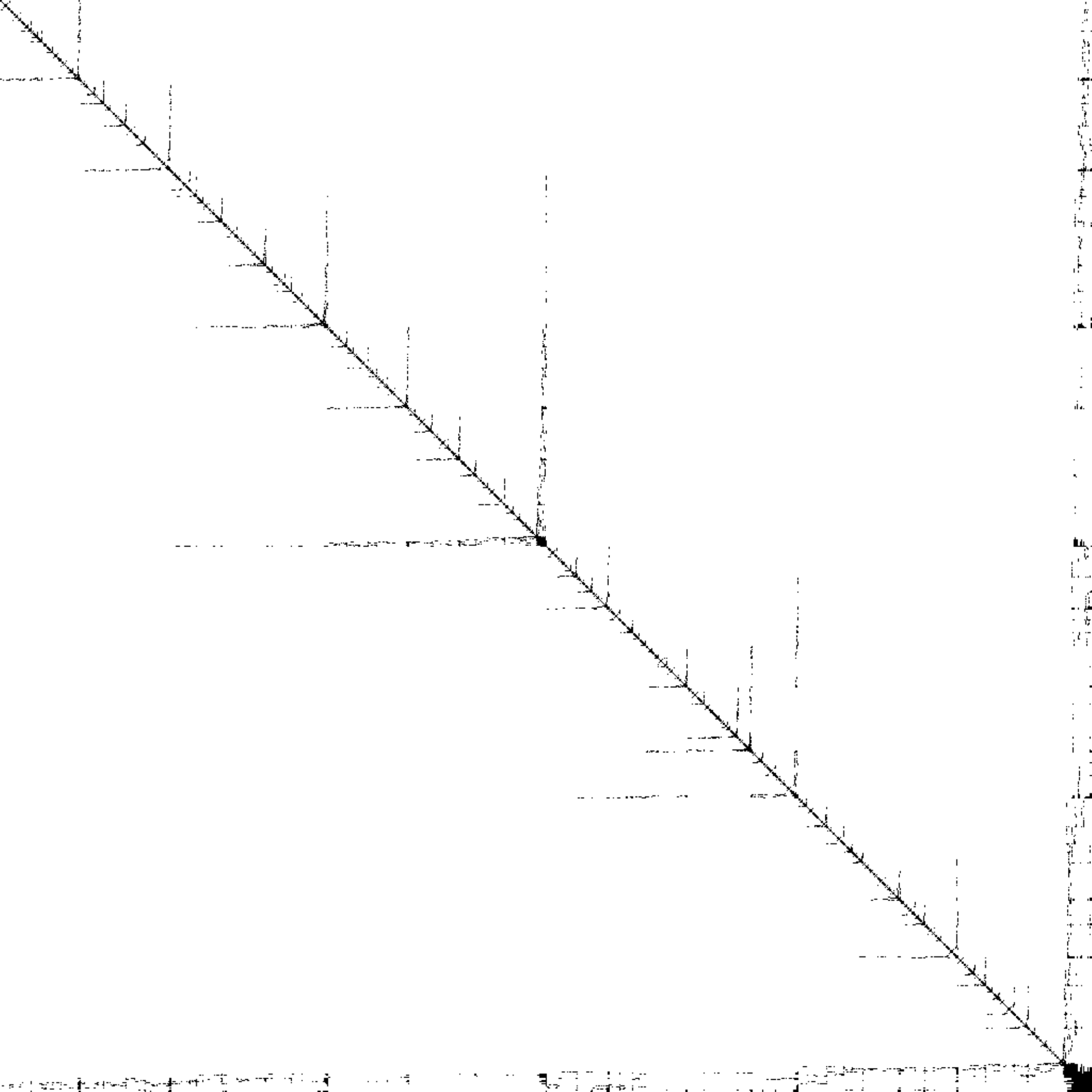}\\
  \caption{Typical distribution of nonzero elements of LU factors.}\label{fig:dist}
\end{figure}

\begin{figure}[t]
  \centering
  \includegraphics[width=1\columnwidth]{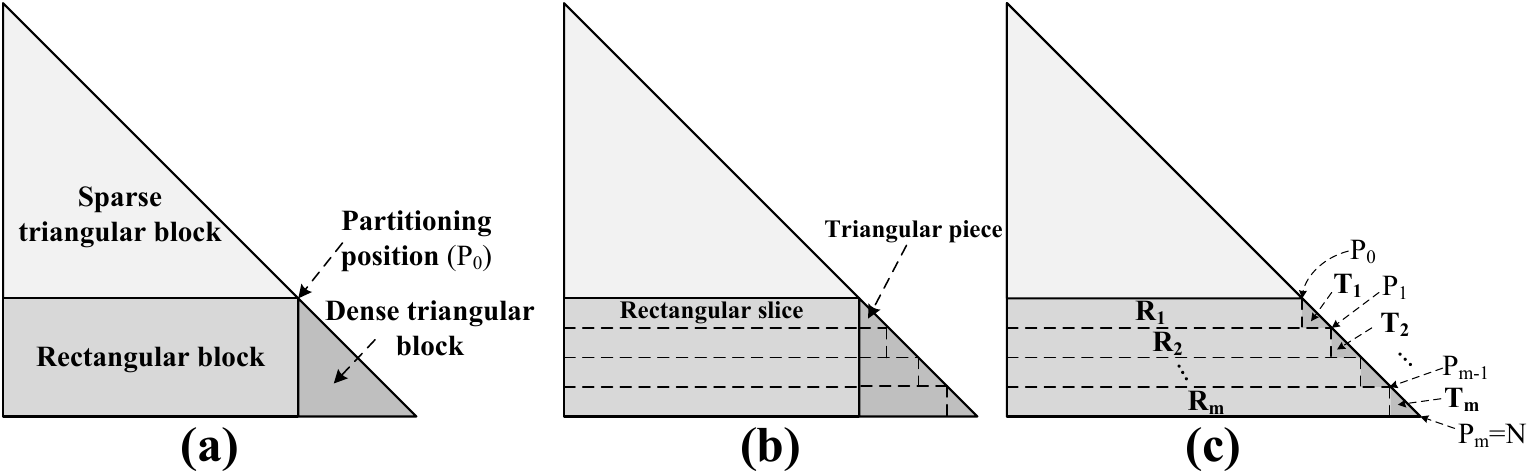}\\
  \caption{Partitioning lower triangular matrix. (a) Coarse-grained partitioning. (b) Fine-grained partitioning. (c) Final partitioning.}\label{fig:partition}
\end{figure}

\subsection{Structure-Adaptive Triangular Matrix Partitioning}\label{sec:partition}
To address the challenges of low parallelism and low computational workload of triangular solving, CKTSO employs a structure-adaptive hybrid parallel  triangular solving method, which is based on a structure-adaptive triangular matrix partitioning method. Fig.~\ref{fig:dist} draws a typical distribution of the nonzero elements of the LU factors. As can be seen, the majority of the nonzero elements are located at the right-bottom corner, while there are two boundaries at the bottom and right sides, respectively. For large-scale circuit matrices, the size of the dense block at the right-bottom corner is usually several hundred to several thousand. Consequently, using the lower triangular matrix as an example, it can be partitioned into three blocks like Fig.~\ref{fig:partition}(a), which are called a sparse triangular block, a rectangular block, and a dense triangular block, respectively. 
Since different matrices have diverse nonzero element distribution and sparsity, the partitioning position ($P_0$) should be adaptively determined. Considering that the dense triangular block dominates the total solving time, the partitioning position is mainly determined by considering the density of the dense triangular block.

In CKTSO, the  partitioning position ($P_0$) is adaptively determined by the following two rules. First, the dense triangular block should have at least 70\% of the nonzero elements of $\bf L$. Second, the dense triangular block should have at least 300,000 nonzero elements. The two thresholds are empirically determined based on lots of experiments. The two conditions guarantee that the dense triangular block is not too sparse or too small. If it is very small or sparse, its computational workload will be small but the scheduling overhead will dominate the solving time. $P_0$ is selected to be the minimum value such that both conditions can hold. If the two conditions cannot hold simultaneously, that is,  the dense triangular block is too small or sparse, $\bf L$ is not partitioned and it is treated as a single sparse triangular block. The following contents will show how to further partition the rectangular and dense triangular blocks, if the two conditions can hold.

The rectangular block is trivial to parallelize. However, the dense triangular block involves strong dependencies and it is almost impossible to be solved in parallel, because the rows are almost one-by-one dependent. To realize efficient parallel triangular solving, the rectangular block is further partitioned into several slices, and the dense triangular block is also further partitioned into rectangular slices and triangular pieces, as shown in Fig.~\ref{fig:partition}(b). To simplify the scheduling, the rectangular slices in the rectangular block and the triangular block are merged. The final partitioning scheme is shown in Fig.~\ref{fig:partition}(c). Let $m$ be the number of slices/pieces. The rectangular slices are denoted by ${\bf R_1},{\bf R_2},\cdots,{\bf R_m}$, and the triangular pieces are denoted by ${\bf T_1},{\bf T_2},\cdots,{\bf T_m}$. The partitioning positions are denoted by $P_0,P_1,\cdots,P_{m}$, where $P_m=N$.

How to determine $m$ should be discussed. The triangular pieces (${\bf T_1},{\bf T_2},\cdots,{\bf T_m}$) involve strong dependencies so they are solved in sequential. It is well known that the sequential workload will greatly impact the overall parallel efficiency, according to the Amdahl's law~\cite{amdahl}. As can be seen, with larger $m$, the sequential workload of the triangular pieces becomes smaller. However, the relative scheduling overhead becomes higher as the workload of each piece becomes smaller. To balance the benefit and overhead, CKTSO uses $m=8$. The partitioning positions ($P_1,\cdots,P_{m-1}$) are simply determined such that the trapezoid slices (${\bf R_1}|{\bf T_1},{\bf R_2}|{\bf T_2},\cdots,{\bf R_m}|{\bf T_m}$)\footnote{Here, the operation  $|$ means joining two matrices to get a single matrix.} have the same number of nonzero elements.

\begin{algorithm}[t]
\caption{Parallel lower triangular solving of ${\bf Ly}=\bf b$.}\label{algo:psolve}
\KwIn{Matrix $\bf L$ obtained from LU factorization, vector $\bf b$, and partitioning positions $P_0,P_1,\cdots,P_{m-1}$}
\KwOut{Vector $\bf y$}
${\bf y}={\bf b}$\;
\tcc{Parallel cluster mode in sparse triangular block}
\For{level in cluster mode}
{
\For {threads \textit{in parallel}}
{
\For{row $i$ assigned to this thread}
{
Solving kernel (lines 3-6 of Algorithm~\ref{algo:solve})\;
}
}
\textbf{barrier}\;
}
\tcc{Sequential solving for remaining rows in sparse triangular block by thread 0}
\If{this is thread 0}
{
\For{each remaining row $i$ in sparse triangular block}
{
Solving kernel (lines 3-6 of Algorithm~\ref{algo:solve})\;
}
}
\textbf{barrier}\;
\tcc{Trapezoid slices}
\For{$k=1:m$}
{
\tcc{Parallel solving for a rectangular slice, i.e., ${\bf y}(P_{k-1}+1:P_k)-={\bf R_k}\cdot{\bf y}(1:P_{k-1})$}
\For{threads in parallel}
{
\For{row $i$ assigned to this thread}
{
$s=0$\;
\For {$j=1:P_{k-1}$ where ${\bf L}(i,j)$ is nonzero}
{
$s+={\bf y}(j)\times {\bf L}(i,j)$\;
}
${\bf y}(i)-=s$\;
}
}
\textbf{barrier}\;
\tcc{Sequential solving for a triangular piece, i.e., solving ${\bf T_k}{\bf y}(P_{k-1}+1:P_k)={\bf y}(P_{k-1}+1:P_k)$}
\If {this is thread 0}
{
\For {$i=P_{k-1}+1:P_{k}$}
{
$s=0$\;
\For {$j=P_{k-1}+1:i-1$ where ${\bf L}(i,j)$ is nonzero}
{
$s+={\bf y}(j)\times {\bf L}(i,j)$\;
}
${\bf y}(i)=({\bf y}(i)-s)/{\bf L}(i,i)$\;
}
}
\textbf{barrier}\;
}

\end{algorithm}

\subsection{Hybrid Parallel Triangular Solving}

After the triangular matrix is partitioned, different parallel solving methods will be used for different sub-matrices, according to the features of the nonzero element distribution. The design of the parallel triangular solving algorithm should fully explore the parallelism with minimum scheduling overhead. Algorithm~\ref{algo:psolve} shows the parallel lower triangular solving algorithm implemented in CKTSO, while the parallel upper triangular solving uses a similar method.

\subsubsection{Sparse Triangular Block}
The sparse triangular block  is typically the largest part with the fewest nonzero elements. If the two conditions mentioned in Section~\ref{sec:partition} cannot hold simultaneously, the entire lower triangular matrix will be treated as a single sparse triangular block.  The nonzero elements are distributed very dispersedly in this block. As a result, it is difficult to block the nonzero elements. Hence, CKTSO also employs a levelization-based method to solve this block. According to the inter-row dependencies determined by the symbolic structure of the sparse triangular block, a dependency graph can also be built and levelized. Different from parallel numerical factorization which uses hybrid cluster and pipeline modes, when solving the sparse triangular block, CKTSO only parallelizes the rows belonging to the cluster mode, but for the remaining nodes, CKTSO solves them in sequential rather than using the pipeline mode. The reason is explained as follows. If dependent rows are solved by the pipeline mode, in Algorithm~\ref{algo:solve}, when executing line 5: $s+={\bf y}(j)\times {\bf L}(i,j)$ for solving ${\bf y}(i)$, it needs to wait for ${\bf y}(j)$ to finish. This implies that, each MAC operation will need a waiting operation. As a result, the parallel overhead will be much higher than the computational workload. In Algorithm~\ref{algo:psolve}, lines 2-6 correspond to the parallel cluster mode. After each level is completed, a barrier is needed to synchronize all threads. Lines 7-9 solve the remaining rows in the sparse triangular block in sequential. 

\subsubsection{Rectangular Slices ($\bf R_k$'s) and Triangular Pieces ($\bf T_k$'s)}
The rectangular slices are of a moderate density and the triangular pieces are the densest part. Each rectangular slice and its corresponding triangular piece form a trapezoid slice (${\bf R_k}|{\bf T_k}$, $k=1,2,\cdots,m$). The $m$ trapezoid slices are solved one by one (line 11), while each rectangular slice is solved in parallel and each triangular piece is solved in sequential. Each rectangular slice is parallelized in a straightforward way. Since the rows  in a rectangular slice are completely independent, they are evenly assigned to all threads, by averaging the number of nonzero elements in the rectangular slice. The computations corresponding to a rectangular slice are equivalent to a general sparse matrix-vector product operation, i.e., ${\bf y}(P_{k-1}+1:P_k)-={\bf R_k}\cdot{\bf y}(1:P_{k-1})$. Lines 12-17 show the parallel solving part for a rectangular slice. After that, a barrier is needed to synchronize all threads. Then the corresponding triangular piece is solved in sequential, which is exactly a triangular solving, i.e., solving ${\bf T_k}{\bf y}(P_{k-1}+1:P_k)={\bf y}(P_{k-1}+1:P_k)$. Lines 20-24 show the solving process for a triangular piece.

\subsection{Setup of Triangular Solving}
The above described parallel triangular solving algorithm needs some setup operations. The setup mainly contains 4 operations: 1) triangular matrix partitioning, 2) levelization of the dependency graph of the sparse triangular block, 3) thread workload assignment, and 4) trapezoid slice segmentation. It should be noted that the dependency graph of the sparse triangular block needs not to be explicitly created, because the inter-row dependencies are implied in the symbolic structure of  $\bf L$ and $\bf U$. The setup is needed only once for a specific symbolic structure of the LU factors. However, if the symbolic structure of the LU factors changes due to re-pivoting, the setup needs to be re-performed.

Triangular matrix partitioning first determines the partitioning position $P_0$. This is done by traversing from $N-1$ to 1 to check the two conditions mentioned in Section~\ref{sec:partition}. If $P_0$ can be found, the partitioning positions $P_1,P_2,\cdots,P_{m-1}$ are further determined by averaging the nonzero elements from row $P_0+1$ to row $N$. This is simply done by traversing the rows and accumulating the row lengths (i.e., the numbers of nonzero elements of the rows). If the current accumulated sum is equal to or larger than the average value (i.e., the total number of nonzero elements from row $P_0+1$ to row $N$ divided by $m$), the accumulation stops and the currently traversed rows are in a trapezoid slice, and then a new accumulation process will start for the next trapezoid slice. The time complexity of triangular matrix partitioning is $O(N)$.

\begin{figure*}[t]
  \centering
  \includegraphics[width=.97\textwidth]{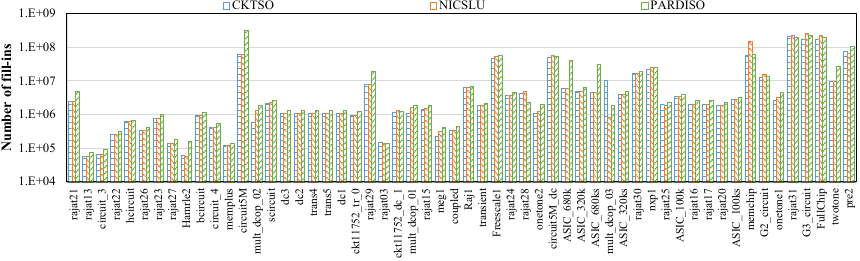}\\
  \caption{Comparison on number of fill-ins.}\label{fig:fillin}
\end{figure*}

\begin{figure*}[!t]
  \centering
  \includegraphics[width=.97\textwidth]{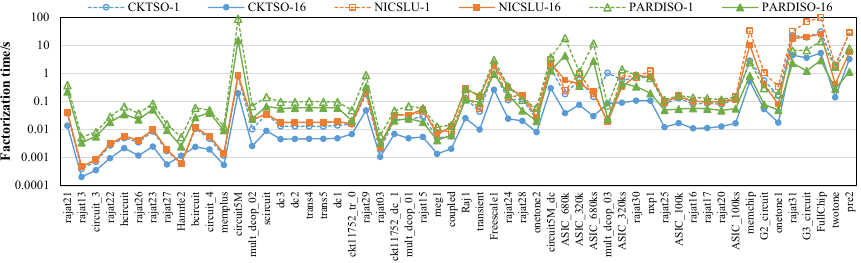}\\
  \caption{Comparison on factorization time (the legend ``solver-$T$'' means the factorization time of ``solver'' with $T$ threads).}\label{fig:ftime}
\end{figure*}

\begin{figure*}[!t]
  \centering
  \includegraphics[width=.97\textwidth]{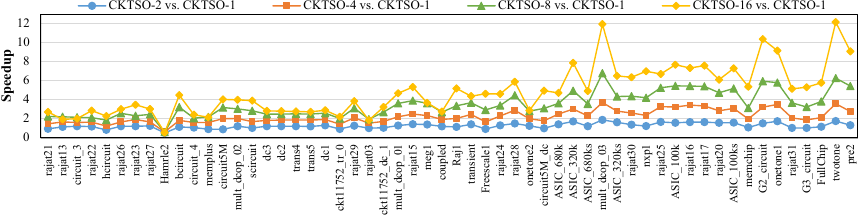}\\
  \caption{Scalability of CKTSO's parallel fast factorization.}\label{fig:fscalability}
\end{figure*}

\begin{figure}[!t]
  \centering
  \includegraphics[width=.8\columnwidth]{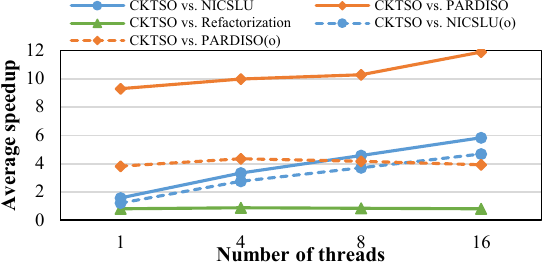}\\
  \caption{Average speedups (across 56 benchmarks) of LU factorization. {NICSLU(o) and PARDISO(o) use the ordering results of CKTSO.}}\label{fig:fspeedup}
\end{figure}

\begin{figure}[!t]
  \centering
  \includegraphics[width=\columnwidth]{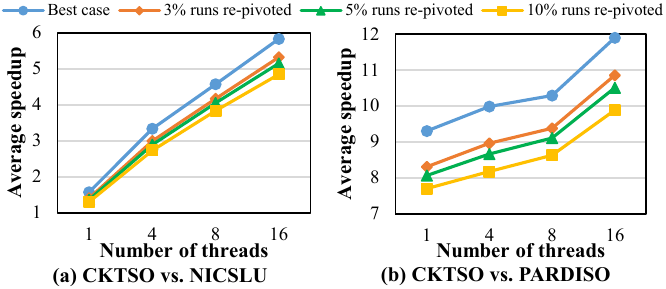}\\
  \caption{{Average speedups (across 56 benchmarks) of fast LU factorization in general cases where some runs incur re-pivoting.}}\label{fig:generalcase}
\end{figure}

Levelization of the dependency graph of the sparse triangular block needs to traverse the nonzero elements of the triangular matrix to calculate the levels of the rows. The time complexity is $O(NNZ)$, where $NNZ$ is the number of nonzero elements in the LU factors. 

Thread workload assignment uses a similar method to the determination of the partitioning positions $P_1,P_2,\cdots,P_{m-1}$, to ensure that each thread processes a similar number of nonzero elements. The time complexity of thread workload assignment is $O(N)$.

Trapezoid slice segmentation can speed up the solving process for the trapezoid slices. The nonzero elements in each row of the trapezoid slices need to be divided into two segments, the rectangular slice and the triangular piece. However, since the column indexes of the LU factors are out of order in each row after the up-looking LU factorization, the segmentation is not trivial.  Completely sorting each row is unnecessary, but for each row in trapezoid slice ${\bf R_k}|{\bf T_k}$, the nonzero elements before and after $P_{k-1}$ need to be separated. CKTSO uses an algorithm that is similar to a single step of quicksort. That is, each row is partitioned into two segments by swapping the elements, according to whether their column indexes are less than or greater than $P_{k-1}$. The time complexity of trapezoid slice segmentation is $O(NNZ)$. The segmentation of different rows can be trivially parallelized.


In summary, the time complexity of the setup of triangular solving is $O(NNZ)$, which equals the time complexity of triangular solving. According to experiments, the setup time is about 2$\times$ of a solving time. Considering that  the symbolic structure of the LU factors can be reused in several tens or more NR iterations in practical SPICE simulations, the setup overhead is negligible.

\section{Experimental Results}
CKTSO is compared with NICSLU~\cite{tcad13,tcasii} and Intel oneMKL PARDISO~\cite{pardiso}. 
 NICSLU is a state-of-the-art sparse solver specially designed for circuit
simulations, which demonstrates outstanding performance in real circuit and power grid simulation problems~\cite{power_pes2015,power_2019,power2015}, while PARDISO
is a general-purpose sparse solver and generally performs
the best among a number of popular sparse solvers for non-circuit
matrices~\cite{compare}.  Experiments are run on a Linux server equipped with an Intel Xeon Gold 6130 CPU (16 cores, 2.1-3.7GHz, 22MB last-level cache) and 256GB memory.  Fifty-six circuit matrices (dimensions from 2,904 to
5,558,326) from the SuiteSparse Matrix Collection~\cite{davis_matrix_web} are tested. {Unless otherwise stated, most benchmark tests correspond to the best case of CKTSO's fast factorization, that is, no re-pivoting happens during pivot check-based fast factorization. General cases that incur re-pivoting are also tested.} CKTSO is also compared with NICSLU in an in-house SPICE simulator for OP and transient simulations.

The 56 benchmarks will be shown in the order of \textit{arithmetic density}, 
which is defined to be $\frac{\#FLOPs}{NNZ({\bf L}+{\bf U}-{\bf I})}$, where $\#FLOPs$ is the number of floating-point operations (FLOPs) involved in numerical LU factorization.

\subsection{Benchmark Tests}

\subsubsection{Number of Fill-ins}
Fig.~\ref{fig:fillin} compares the number of fill-ins among CKTSO, NICSLU, and PARDISO. NICSLU  uses the approximate minimum degree algorithm~\cite{amd} and its variants~\cite{mf}, while PARDISO uses nested dissection implemented by METIS~\cite{metis}. On average, NICSLU and PARDISO generate 8.8\% and 62.3\% more fill-ins than CKTSO, respectively. Minimum degree performs well for most benchmarks. Only for a few large-scale, post-layout, or mesh-style benchmarks, such as G3\_circuit, G2\_circuit, memchip, meg1, and FullChip, minimum degree generates significantly more fill-ins than nested dissection. Nested dissection adopted by PARDISO generates more fill-ins than minimum degree for most benchmarks. 
On average, NICSLU and PARDISO generate 39.8\% and 188\% 
 more FLOPs than CKTSO, respectively.  This comparison reveals that CKTSO generally obtains the fewest fill-ins for a wide range of circuit matrices, by combing minimum degree and nested dissection methods.

\subsubsection{Performance of Fast Factorization}

Fig.~\ref{fig:ftime} compares CKTSO's fast factorization with NICSLU's and PARDISO's factorization. It can be observed that CKTSO generally has the smallest factorization time.  CKTSO is the fastest for 50 out of the 56 benchmarks, among the compared solvers. Due to the reduced fill-ins and FLOPs brought by the matrix ordering method, sequential CKTSO is even faster than parallel NICSLU and PARDISO for half of the benchmarks.

Fig.~\ref{fig:fscalability} shows the scalability of CKTSO's fast factorization. For the average speedup across the 56 benchmarks, the fast factorization is sped up by 1.27$\times$, 2.23$\times$, 3.5$\times$, and 4.86$\times$ when using 2, 4, 8, and 16 threads, respectively. The scalability tends to be better for matrices with higher arithmetic density. This is easy to understand, since with higher arithmetic density, the relative computational workload is larger and the parallel scheduling overhead is relatively smaller.

When compared with parallel factorization of NICSLU and PARDISO, CKTSO's fast factorization achieves good speedups, as shown in Fig.~\ref{fig:fspeedup}. CKTSO is consistently more than 9$\times$ faster than PARDISO on average, when both solvers use 1, 4, 8, and 16 threads, respectively. CKTSO's parallel fast factorization shows better scalability than both NICSLU's and PARDISO's parallel factorization, as the average speedups increase with parallelism increasing. When the solvers all use 16 threads, CKTSO is on average 5.9$\times$ and 11.9$\times$ faster than NICSLU and PARDISO, respectively. CKTSO's parallel fast factorization is also compared with CKTSO's parallel re-factorization, and the average speedups are consistently about 0.8. This implies that the proposed fast factorization has similar scalability to re-factorization. {To eliminate the impact of matrix ordering, Fig.~\ref{fig:fspeedup} also compares CKTSO with NICSLU and PARDISO when the latter two use the ordering results of CKTSO, denoted as NICSLU(o) and PARDISO(o). In this case, CKTSO is on average 1.21$\times$, 2.74$\times$, 3.71$\times$, and 4.69$\times$ faster than NICSLU(o) when using 1, 4, 8, and 16 threads, respectively. The performance of PARDISO(o) is improved by 2-3$\times$ when using CKTSO's ordering, but CKTSO is still about 4$\times$ faster than PARDISO(o) on average.}

{To evaluate the general case of fast factorization, each matrix is factorized 100 times where 3\%, 5\%, and 10\% runs are randomly selected to perform re-pivoting.  If a run is selected to perform re-pivoting, the re-pivoting row is randomly selected from $[1,N]$. The average factorization time of 100 runs is evaluated as
the general-case performance and compared with the performance
of NICSLU and PARDISO, as shown in Fig.~\ref{fig:generalcase}. The increase of the number of runs that incur re-pivoting slightly decreases the speedups. However, even with 10\% runs incurring re-pivoting, the speedups are dropped only by about 20\%.}

\subsubsection{Performance of Triangular Solving}

\begin{figure*}[!t]
  \centering
  \includegraphics[width=.97\textwidth]{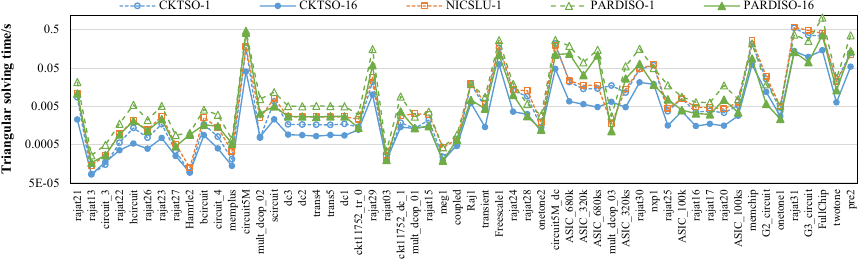}\\
  \caption{Comparison on triangular solving time (the legend ``solver-$T$'' means the factorization time of ``solver'' with $T$ threads).}\label{fig:stime}
\end{figure*}

\begin{figure*}[!t]
  \centering
  \includegraphics[width=.97\textwidth]{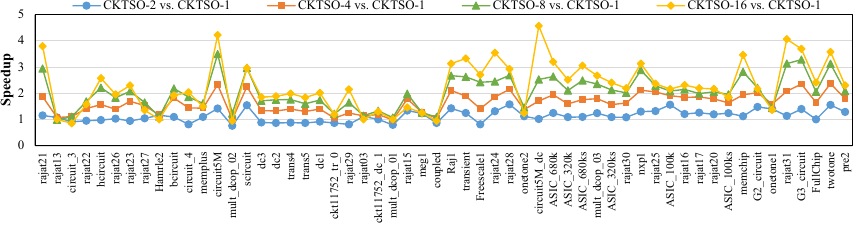}\\
  \caption{Scalability of CKTSO's parallel triangular solving.}\label{fig:sscalability}
\end{figure*}

\begin{figure}[!t]
  \centering
  \includegraphics[width=.8\columnwidth]{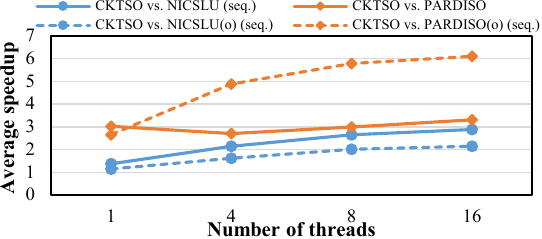}\\
  \caption{Average speedups (across 56 benchmarks) of triangular solving. {NICSLU(o) and PARDISO(o) use the ordering results of CKTSO.}}\label{fig:sspeedup}
\end{figure}

Fig.~\ref{fig:stime} compares the triangular solving time. NICSLU only has sequential solving. Like LU factorization, CKTSO also generally has the smallest solving time.  CKTSO is the fastest for 49 out of the 56 benchmarks, among the compared solvers.

Fig.~\ref{fig:sscalability} shows the scalability of CKTSO's parallel solving.  On average, CKTSO's parallel solving achieves 1.2$\times$, 1.63$\times$, 2.04$\times$, and 2.3$\times$ speedups when using 2, 4, 8, and 16 threads, respectively. The scalability of solving is lower than that of LU factorization, due to the smaller computational workload and lower parallelism of triangular solving.

As shown in Fig.~\ref{fig:sspeedup}, when compared with NICSLU's sequential solving, CKTSO's parallel solving achieves 1.37-2.89$\times$ average speedups when CKTSO uses 1-16 threads. When compared with PARDISO, CKTSO's solving consistently shows high speedups (2.7-3.3$\times$), when both solvers use 1-16 threads. {When NICSLU and PARDISO use CKTSO's ordering results, CKTSO's solving is 1.15-2.15$\times$ faster than NICSLU(o)'s. PARDISO does not support parallel solving when using external ordering and CKTSO's solving is 2.65-6.1$\times$ faster than PARDISO's on average.}

\subsection{SPICE Simulation Tests}

To evaluate the performance of CKTSO in real SPICE simulations, CKTSO and NICSLU are integrated in an in-house SPICE simulator, respectively. The parallel fast factorization is evaluated in OP simulations of nonlinear circuits, while the parallel triangular solving is evaluated in transient simulations of linear circuits.

Table~\ref{tab:dc} compares the OP simulation time of 5 nonlinear circuits. Both CKTSO and NICSLU use 16 threads for factorization and 1 thread for triangular solving. CKTSO achieves an average speedup of 2.38$\times$ compared with NICSLU in total OP simulation time. Note that besides the sparse solver, there are other core operations (e.g., transistor model evaluation) in a SPICE simulation which cost about half time, so the speedups of the OP simulation time are lower than those of pure LU factorization which are shown in Fig.~\ref{fig:fspeedup}. As expected, for most NR iterations, CKTSO computes the entire matrix without re-pivoting, which is the main motivation of designing CKTSO. There are indeed some iterations incurring re-pivoting. {Generally, re-pivoting happens when the difference between the solutions of two successive NR iterations is large. This usually implies a sharp change in the convergence path of the NR method which is typically caused by significant changes in the matrix values.} This explains the necessity of using the proposed fast factorization algorithm to replace pure re-factorization without pivoting, as the latter may generate inaccurate solutions in those iterations that need re-pivoting.

Table~\ref{tab:tran} compares the transient simulation time of 6 ibmpg circuits~\cite{ibmpg}. Each circuit simulation executes one LU factorization and 1000 triangular solvings (1000 time nodes with a fixed time step). Triangular solving spends about 3/4 of the total simulation time for these cases. CKTSO-1 and CKTSO-16 achieve average speedups of 1.27$\times$ and 2.56$\times$ compared with NICSLU-1 in transient simulations, respectively. CKTSO-16 is on average 1.97$\times$ faster than CKTSO-1 in transient simulations.

\begin{table}[t]
\centering
\begin{threeparttable}
\caption{Comparison on SPICE OP simulation time of nonlinear circuits.}
\label{tab:dc}
\begin{tabular}{cc|cc|ccc}
\toprule
\multirow{2}{*}{\#Nodes\tnote{a}}&\multirow{2}{*}{Dim.\tnote{b}}&\multicolumn{2}{c|}{NICSLU-16}&\multicolumn{3}{c}{CKTSO-16}\\
{}&{}&{Time/s}&{\#Iter.}&{Time/s}&{\#Iter.}&{\#Re-piv.\tnote{c}}\\
\midrule
{11705}&{12088}&{2.196}&{89}&{0.765}&{89}&{3}\\
{26564}&{46820}&{6.098}&{236}&{5.722}&{236}&{2}\\
{63013}&{63929}&{22.36}&{133}&{9.029}&{133}&{3}\\
{341965}&{344285}&{470.4}&{305}&{197.4}&{305}&{14}\\
{564985}&{567895}&{8722}&{1939}&{2810}&{1997}&{26}\\
\bottomrule
\end{tabular}
 \begin{tablenotes}
 \item[a] Number of nodes in the circuit, which is reported by the simulator.
 \item[b] Dimension of the created matrix.
  \item[c] Number of iterations in which re-pivoting happens.
  \end{tablenotes}
\end{threeparttable}
\end{table}

\begin{table}[t]
\centering
\caption{Comparison on SPICE transient simulation time of linear circuits.}
\label{tab:tran}
\setlength{\tabcolsep}{5pt}
\begin{tabular}{ccc|c|cc}
\toprule
\multirow{2}{*}{Circuit} & \multirow{2}{*}{\#Nodes} &\multirow{2}{*}{Dim.} &NICSLU-1 &CKTSO-1 &CKTSO-16\\
{}&{}&{}&{Time/s}&{Time/s}&{Time/s}\\
\midrule
{ibmpg1t}&{39681}&{53988}&{1.56}&{1.56}&{1.42}\\
{ibmpg2t}&{164238}&{164567}&{15.79}&{14.36}&{8.49}\\
{ibmpg3t}&{1041535}&{1042489}&{187}&{125.5}&{53.9}\\
{ibmpg4t}&{1212365}&{1213326}&{224.3}&{148.4}&{67.9}\\
{ibmpg5t}&{1552785}&{2091871}&{212.5}&{167.5}&{76.4}\\
{ibmpg6t}&{2367183}&{3203421}&{290.4}&{236}&{101.6}\\
\bottomrule
\end{tabular}
\end{table}

\section{Conclusions}

Conventionally, in many practical applications, sparse solvers are usually made as black boxes  and used as standalone modules. However, if the special features of the matrices involved in the targeted application are taken into account, the solver design can be more dedicated for elevating performance. This paper introduces CKTSO, a parallel sparse linear solver specially optimized for SPICE simulators. By considering the features of the matrix sparsity and the slow change in matrix values in SPICE simulations, the three steps of a sparse direct solver, pre-processing, numerical factorization, and triangular solving, have been elaborated. Both benchmark tests and SPICE simulation tests have revealed the superior performance of CKTSO, in comparison with state-of-the-art solvers.

\bibliographystyle{IEEEtran}
\bibliography{ref}

\begin{IEEEbiography}[{\includegraphics[width=1in,height=1.25in,clip,keepaspectratio]
{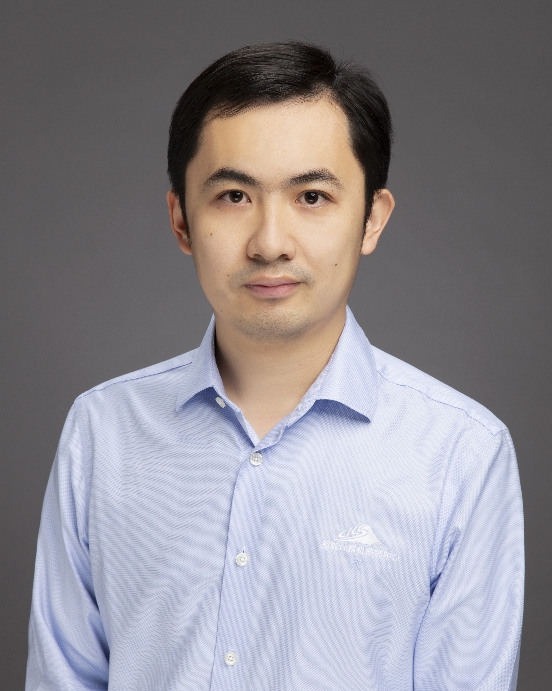}}]{Xiaoming Chen} (S'12-M'15) received the BS and PhD degrees in electronic engineering from Tsinghua University, Beijing, China, in 2009 and 2014, respectively.

He is a Professor with Institute of Computing Technology, Chinese Academy of Sciences. His research interests include design automation for integrated circuits and computer architectures. He has published 1 Springer book and about 130 papers in DAC, ICCAD, DATE, MICRO, HPCA, ASPLOS, IEEE TCAD, IEEE TC, IEEE TPDS, etc. His researches have been adopted by commercial EDA software.  He was a recipient of the NSFC Excellent Young Scientists Fund, 2016 European Design and Automation Association (EDAA) Outstanding Dissertation Award, 2018 DAMO Academy Young Fellow Award, 2023 Hot Paper Award of SCIENCE CHINA Information Sciences, and ASP-DAC 2022 Best Paper Award.
\end{IEEEbiography}

\end{document}